\documentclass[preprint]{aastex}

\usepackage{color}

\def\araa{ARA\&A}%
\def\apj{ApJ}%
\def\apjl{ApJ}%
\def\apjs{ApJS}%
\def\aap{A\&A}%
\def\mnras{MNRAS}%
\def\prc{Phys.~Rev.~C}%
\def\gca{Geochim.~Cosmochim.~Acta}%
\def\nar{New~Astr.~Rev.}%

\include{abbrev}
\include{abbr-engl}

\slugcomment{DRAFT: \today}

\shorttitle{He shell in massive stars}
\shortauthors{}

\begin{document}

\title{Production of carbon-rich presolar grains from massive stars.}
\author{M. Pignatari\altaffilmark{1,10},
M. Wiescher\altaffilmark{2,3},
F.X. Timmes\altaffilmark{3,4,10},
R.J. de Boer\altaffilmark{2,3},
F.-K. Thielemann\altaffilmark{1},
C. Fryer\altaffilmark{5,10},
A. Heger\altaffilmark{6,10},
F. Herwig\altaffilmark{3,7,10},
R. Hirschi\altaffilmark{8,9,10}}
\altaffiltext{1}{Department of Physics, University of Basel, Klingelbergstrasse 82, CH-4056 Basel, Switzerland}
\altaffiltext{2}{University of Notre Dame, Department of Physics, Notre Dame, Indiana 46556, USA}
\altaffiltext{3}{The Joint Institute for Nuclear Astrophysics, Notre Dame, IN 46556, USA}
\altaffiltext{4}{Arizona State University, School of Earth and Space Exploration, PO Box 871404, Tempe, AZ, 85287-1404, USA.}
\altaffiltext{5}{Computational Physics and Methods (CCS-2), LANL, Los Alamos, NM, 87545, USA.}
\altaffiltext{6}{Monash Centre for Astrophysics, School of Mathematical Sciences, Monash University, Vic 3800, Australia.}
\altaffiltext{7}{Department of Physics \& Astronomy, University of Victoria, Victoria, BC,
\altaffiltext{8}{Keele University, Keele, Staffordshire ST5 5BG, UK.}
\altaffiltext{9}{Institute for the Physics and Mathematics of the Universe, University of Tokyo, 5-1-5 Kashiwanoha, Kashiwa 277-8583, Japan}
V8P5C2 Canada.}
\altaffiltext{10}{NuGrid collaboration,  \url{http://www.nugridstars.org}}

\begin{abstract}
About a year after core collapse supernova, dust starts to condense
in the ejecta. In meteorites, a fraction of C-rich presolar grains
(e.g., silicon carbide (SiC) grains of Type-X and low density graphites)
are identified as relics of these events,
according to the anomalous isotopic abundances.
Several features of these abundances remain unexplained and
challenge the understanding of core-collapse supernovae explosions and
nucleosynthesis.
We show, for the first time, that most of the measured C-rich grain
abundances can be accounted for in the C-rich material from explosive
He burning in core-collapse supernovae
with high shock velocities and consequent high temperatures.
The inefficiency of the $^{12}$C($\alpha$,$\gamma$)$^{16}$O reaction
relative to the rest of the $\alpha$-capture chain at $T >
3.5\times10^8 \mathrm{K}$ causes the deepest He-shell material to be carbon rich and
silicon rich, and depleted in oxygen. The isotopic ratio predictions
in part of this material, defined here as the C/Si zone,
are in agreement with the grain data.
The high-temperature explosive conditions that our models reach at the bottom of
the He shell, can also be representative of the nucleosynthesis
in hypernovae or in the high-temperature tail of a
distribution of conditions in asymmetric supernovae.
Finally, our predictions are consistent with the observation of large
$^{44}$Ca/$^{40}$Ca observed in the grains. This is due to the production
of $^{44}$Ti together with $^{40}$Ca in the C/Si zone, and/or to the strong
depletion of $^{40}$Ca by neutron captures.
\end{abstract}

\section{Introduction}
\label{sec:intro}

Massive stars (M $\gtrsim$ $8 \mathrm{M_{\odot}}$) are among the major
contributors to the chemical enrichment of the galaxy
through both their winds and core collapse Supernova (CCSN).
In carbonaceous meteorites several types of
C-rich presolar grains condensed in CCSN ejecta have been
identified through association of the measured isotopic abundance
patterns with signatures of such events
\citep[see e.g.,][]{zinner:03,clayton:04}.
Among those, C-rich grains
like Type X silicon carbides (SiC-X) and graphites
are assumed to require a C-rich environment to form
\citep[C/O $>$ 1, e.g.,][]{travaglio:99}.
Standard massive star models
predict that C-rich ejecta originate in
explosive He shell burning, characterized by partial pre-explosive
He-burning products \citep[e.g.,][]{woosley:95}.
The heavy species abundances show a mild $s$-process signature
 from pre-explosive convective
He-shell burning
\citep[triggered by the $^{22}$Ne($\alpha$,n)$^{25}$Mg reaction,
e.g.,][]{rauscher:02}, and by the explosive neutron burst triggered
again by the $^{22}$Ne($\alpha$,n)$^{25}$Mg reaction
\citep[namely the $n$-process, with n$_{\rm n}$ $\sim$ 10$^{18}$
cm$^{-3}$, e.g.,][]{blake:76,meyer:00}.

%
SiC-X grains show a $^{28}$Si excess compared to
$^{29,30}$Si, and
most of these grains have a $^{12}$C and $^{15}$N excess while some
show a $^{13}$C and $^{15}$N excess.
The $^{26}$Mg and $^{44}$Ca excess compared to solar has been
explained by the condensation of unstable $^{26}$Al
and $^{44}$Ti into the grains
\citep[e.g.,][]{amari:92, nittler:02, besmehn:03}.
Graphite grains \citep[with low density, LD group,][]{amari:95}
show a large range of $^{12}$C/$^{13}$C ratio, possibly correlated
with $^{18}$O excess.
Both excess and deficit of $^{28}$Si with respect to $^{29,30}$Si
can be observed.
Indication of $^{26}$Mg and
$^{44}$Ca excess is also found for several of those grains.
Finally, their measured $^{14}$N/$^{15}$N
ratio with few exceptions is quite normal,
pointing to a contamination from solar nitrogen
\citep[e.g.,][and references therein]{zinner:03}.

To explain these abundance signatures \cite{travaglio:99} and
\cite{yoshida:06} assume mixing of He-shell matter with Si/S zone
material \citep[see][]{meyer:95} deeper in the supernova without any
mixing contribution from the O-rich layers in between.
The main constraint is that the material out of which the grains form
needs to be C-rich, to have free carbon
to condense in SiC and graphites after CO formation.
\cite{clayton:13} and references therein
suggests instead that since SN ejecta are not
in thermodynamic equilibrium, C-rich grains may condense in O-rich and
$^{28}$Si-rich ejecta.

More recently, \cite{marhas:08} found a deficit or
normal $^{54}$Fe compared to $^{56}$Fe abundance in SiC-X grains,
in disagreement with nucleosynthesis predictions for the
$^{28}$Si-rich Si/S zone to be $^{54}$Fe-rich
\citep[e.g.,][]{thielemann:96}.
Possible fractionation
between Si and Fe proposed by \cite{marhas:08} is unlikely to be
sufficient, since Fe-rich subgrains in the SiC-X grains
also show a normal $^{54}$Fe/$^{56}$Fe ratio.

In this Letter we propose a new nucleosynthesis scenario,
where the observational signatures of C-rich presolar grains can
be obtained, without mixing, from the C-rich material in
CCSN that experience a strong shock.
The paper is organized as follows. In \S \ref{sec: models_description}
we describe stellar models and the nucleosynthesis calculations,
in \S \ref{sec: comparison} we compare theoretical results with
measurements for SiC X grains and graphite grains.
Finally in \S \ref{sec: summary} results are summarized.

\section{Stellar model calculations and nucleosynthesis}
\label{sec: models_description}

In this study, we consider three SN explosion models for a
$15 \mathrm{M_{\odot}}$, $\mathrm{Z} = 0.02$ star.
(Pignatari et al. 2013, in preparation).
The pre-supernova evolution is computed with the
stellar code GENEC
\citep[see for more details][model 15ST in their Fig.\,6]{bennett:12}.
The explosion simulations include the fallback prescription by
\cite{fryer:12}, using the recommended initial shock velocity
and reduced by a factor of 2 and 4, respectively
(models 15r, 15r2 and 15r4).
The standard initial shock velocity used beyond fallback is
$2\times10^{9} \mathrm{cm}\mathrm{s}^{-1}$.
For a 15\,M$_\odot$ star simulation by \cite{fryer:12},
a $2 \times 10^9 {\rm cm s^{-1}}$ shock velocity corresponds
to a $3-5 \times 10^{51} {\rm erg}$ explosion, a  $1 \times 10^9 {\rm cm s^{-1}}$
shock velocity corresponds to a $1-2 \times 10^{51} {\rm erg}$ explosion, and the
 $5 \times 10^8 {\rm cm s^{-1}}$ shock velocity corresponds to a weak explosion below
 $10^{51} {\rm erg}$.

The full nucleosynthesis is calculated
using the post-processing code MPPNP
\citep[see e.g.,][]{bennett:12}.
Here we focus only on C-rich
explosive He burning ejecta which include the He/C zone and
a small part of the O/C zone \citep[][]{meyer:95}.

Compared to standard symmetric CCSN models with
explosion energy of 10$^{51}$ erg \citep[e.g., ][]{woosley:95},
our models show that depending on the
explosion energy $^{16}$O is depleted and $^{28}$Si is accumulated
at the bottom of the He/C zone and at the top of
the C/O zone, where $^{12}$C is not significantly modified
(see Fig.\,\ref{fig:15_spagh_he} for the details of models
15r and 15r4 with the largest and smallest shock).
Such $^{28}$Si-enrichment is obtained also in the intermediate model
15r2. Similar nucleosynthesis signatures and high temperature conditions
are also predicted in high energy ejecta of aspherical SN and in
Hypernovae \citep[e.g.,][]{nomoto:09}.

\begin{figure}
\centering
\resizebox{8.3cm}{!}{\rotatebox{0}{\includegraphics{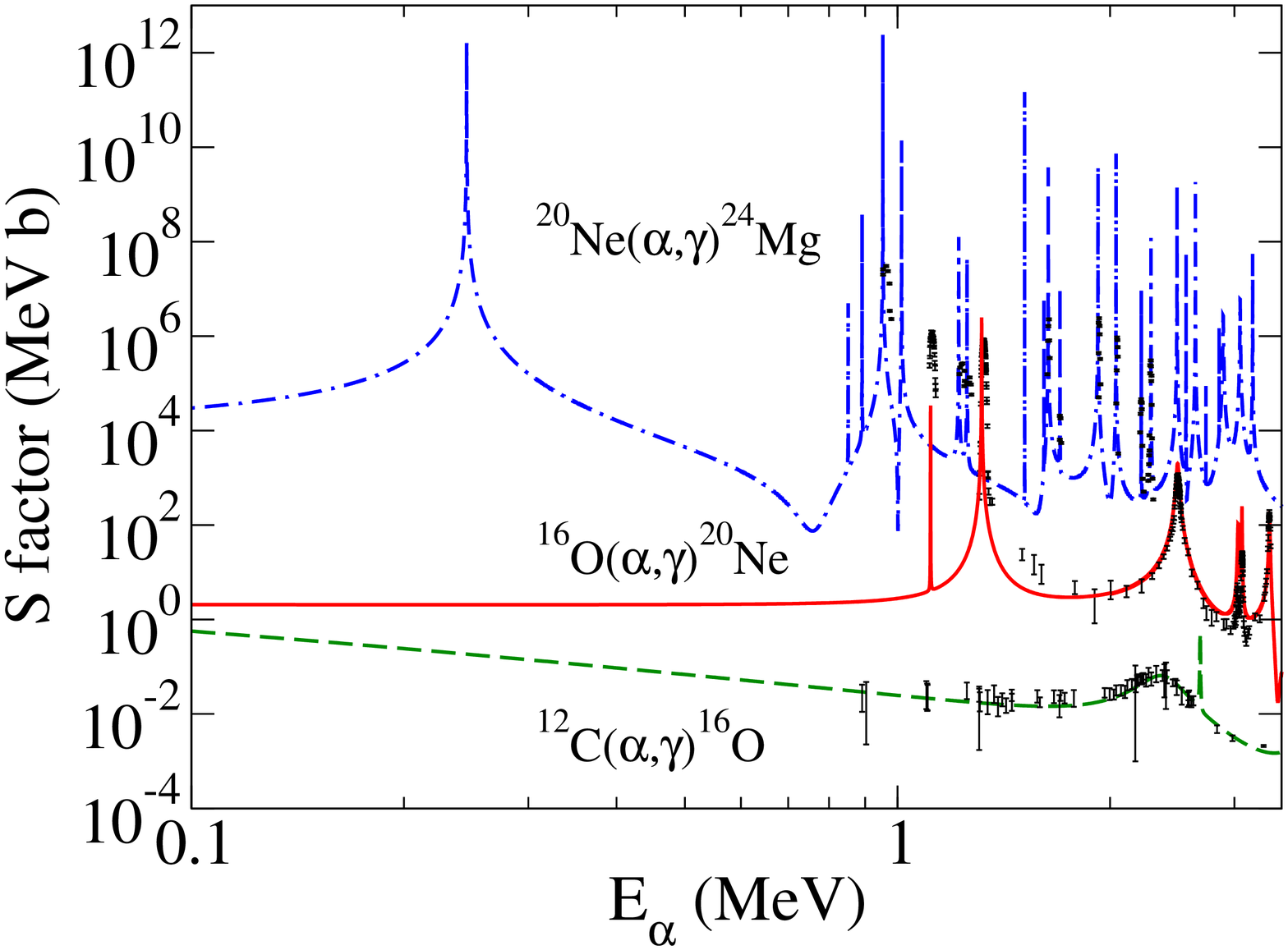}}}
\resizebox{7.5cm}{!}{\rotatebox{0}{\includegraphics{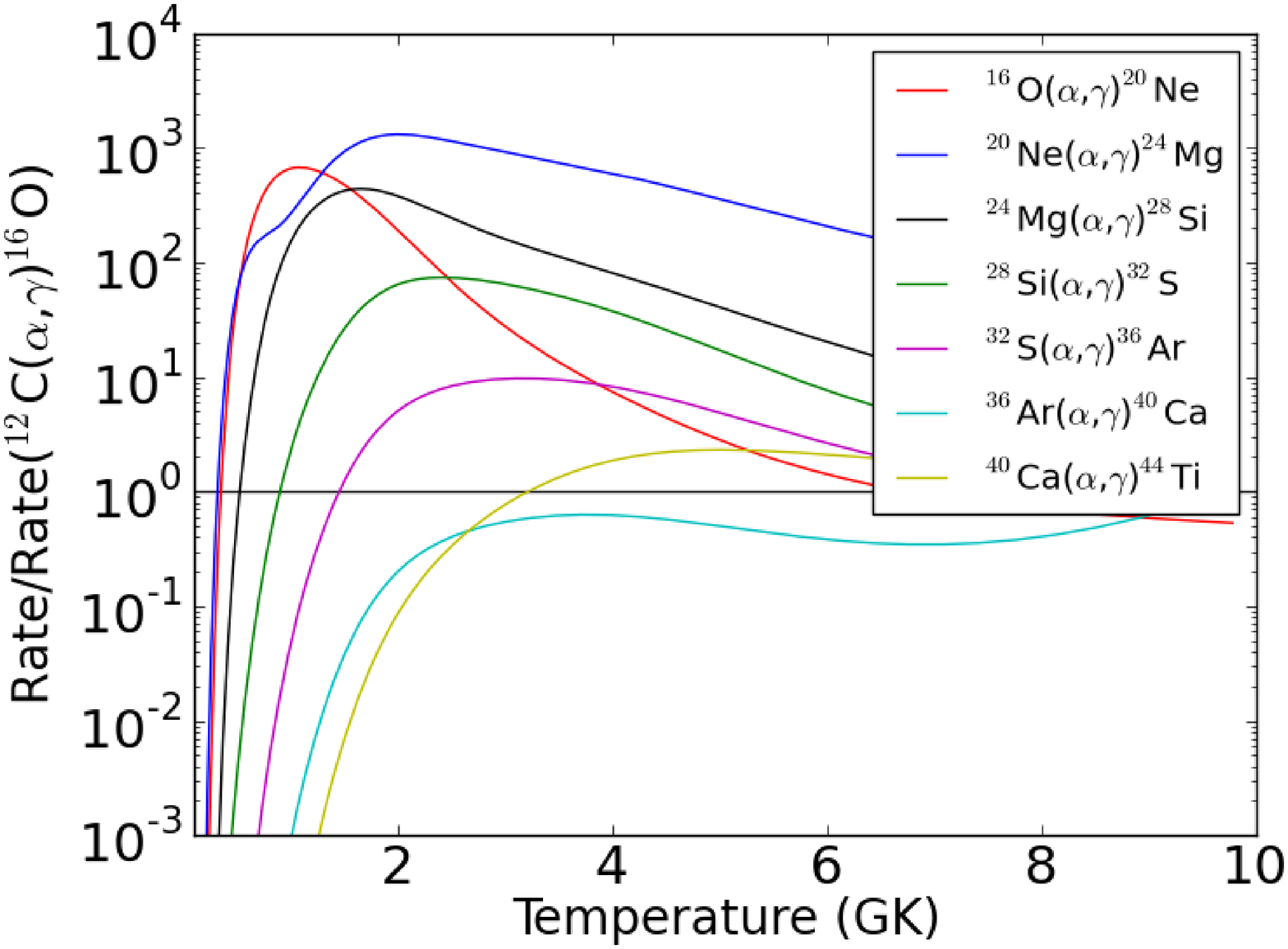}}}
\caption{
The S-factors for the $^{12}$C($\alpha$,$\gamma$)$^{16}$O,
$^{16}$O($\alpha$,$\gamma$)$^{20}$Ne
and $^{20}$Ne($\alpha$,$\gamma$)$^{24}$Mg are shown as function
of $\alpha$-energy in the laboratory system (Left Panel).
The $\alpha$-capture rates of the
$\alpha$-chain from $^{16}$O to $^{44}$Ti and normalized to the
rate of $^{12}$C($\alpha$,$\gamma$)$^{16}$O
are shown as a function of temperature (Right Panel).
}
\label{fig:rates_comp}
\end{figure}

At temperatures above $\sim$ $3.5\times10^8 \mathrm{K}$ the $(\alpha,\gamma)$-rates of
$^{16}$O and $^{20}$Ne exceed the
 $^{12}$C$(\alpha,\gamma)$-rate (Fig.\,\ref{fig:rates_comp}).
The reaction rate is directly correlated with the S-factor.
The rate is obtained by integrating the S-factor over the Gamow window,
the energy range of stellar burning at a certain temperature as defined
by \cite{caughlan:85}.
The S-factor curve of $^{12}$C($\alpha$,$\gamma$)$^{16}$O is
characterized by non-resonant reaction components:
they primarily consist of interfering E1 transitions from tails
of broad high energy 1$^-$ resonances and the 1$^-$ sub-threshold
state, and a E2 direct capture component interfering with
low energy tails of 2$^+$ resonances and the subthreshold state
\citep[][]{buchmann:06}.
While at low temperatures the reaction rate is lower than
$^{16}$O($\alpha$,$\gamma$)$^{20}$Ne, it declines towards
higher temperatures because of the decline of the S-factor
by two orders of magnitude. The $^{16}$O($\alpha$,$\gamma$)$^{20}$Ne
reaction rate on the other hand is determined by a constant non-resonant
S-factor term and by the narrow 3$^-$ and 1$^-$ resonances
at 1.1 MeV and 1.3 MeV respectively, which dominate the reaction rate
above temperatures of $T \sim 10^9 \mathrm{K}$ \citep[][]{costantini:10}.
Concerning the reaction rates of the subsequent
$\alpha$-capture reactions $^{20}$Ne($\alpha$,$\gamma$)$^{24}$Mg
\citep[][]{schmalbrock:83}
and $^{24}$Mg($\alpha$,$\gamma$)$^{28}$Si \citep[][]{strandberg:08},
the non-resonant S-factor terms are
larger and the critical energy range above 0.3 MeV is
characterized by narrow resonances which cause a
further enhancement in the reaction rate, as demonstrated for
$^{20}$Ne($\alpha$,$\gamma$)$^{24}$Mg in Fig.\,\ref{fig:rates_comp}.
This is also true but for larger energies and to a smaller extent for
the production of $^{32}$S, $^{36}$Ar, $^{40}$Ca, and $^{44}$Ti.

In parts of the He/C and C/O zones, $^{16}$O is thus depleted
and feeds the production of $^{28}$Si. This zone is C and Si rich,
and therefore we call this zone the C/Si zone.
A fundamental condition required to form the C/Si zone is that
sufficient $^4$He fuel is available while high enough temperatures
are reached in the explosive burning.

For model 15r, $^{16}$O and $^{20}$Ne are depleted, whereas
$^{28}$Si and $^{32}$S are produced up to a mass fraction
of $\sim$ 20\% in the C/Si zone, due to reactions along the $\alpha$-capture
chain (Fig.\,\ref{fig:rates_comp}). As shown in Fig.\,\ref{fig:15_spagh_he},
the $\alpha$-capture chain is efficient up to $^{44}$Ti.
At the bottom of the He/C zone results are similar, but
since the pre-explosive $^{16}$O abundance is lower
the production of $^{28}$Si and $^{32}$S
is less efficient. In this zone $^{40}$Ca is depleted via neutron
captures, and unstable $^{44}$Ti is not produced.
In model 15r4 the most produced species in the C/Si zone
are $^{20}$Ne, $^{24}$Mg and $^{28}$Si.
In comparison with the pre-explosive abundances,
the species above $^{32}$S and along the $\alpha$-chain
are not produced efficiently. A similar but milder signature
is obtained again at the bottom of the He/C zone.
Notice that the rapid decrease of $\alpha$-particles in the O/C zones
is causing a weaker O-depletion (and $^{28}$Si-excess).
Before the SN explosion such a zone is radiative, still carrying the signature
of the pre-explosive convective He-burning core.
The He profile under the He shell depends on uncertain physics
assumptions in the models.
Furthermore, $^{4}$He may be mixed into the O/C zone
from the bottom of the He shell by extra-mixing triggered by
processes that have not been considered in our pre-SN model, such as
rotation \citep[e.g.,][]{meynet:06} or convective boundary mixing
\citep[e.g.,][]{meakin:07}.
In that case larger fractions of the O/C zone would
be converted into the C/Si zone compared to our model.

In summary, in the CCSN ejecta exposed to strong shocks
the upper part of the C/O zone and an extended region of
the He/C zone are $^{16}$O-depleted and $^{28}$Si-rich, and
$^{28}$Si will be present in C-rich material when grains
will start to condense.

\section{Comparison with observations}
\label{sec: comparison}

In this section, the nucleosynthesis predictions
described in \S \ref{sec: models_description} are compared
with observations for single SiC-X and graphite grains,
from the St.\, Louis Presolar Grains Database
\citep[][]{hynes:09}.
We compare $only$ the abundances from the C-rich
region in the He shell (namely, the C/Si zone and the He/C zone).
Mildly C-rich regions in the He/N zone (H-burning layer)
will be discussed in a future work.

\begin{figure}
\centering
\resizebox{7.5cm}{!}{\rotatebox{0}{\includegraphics{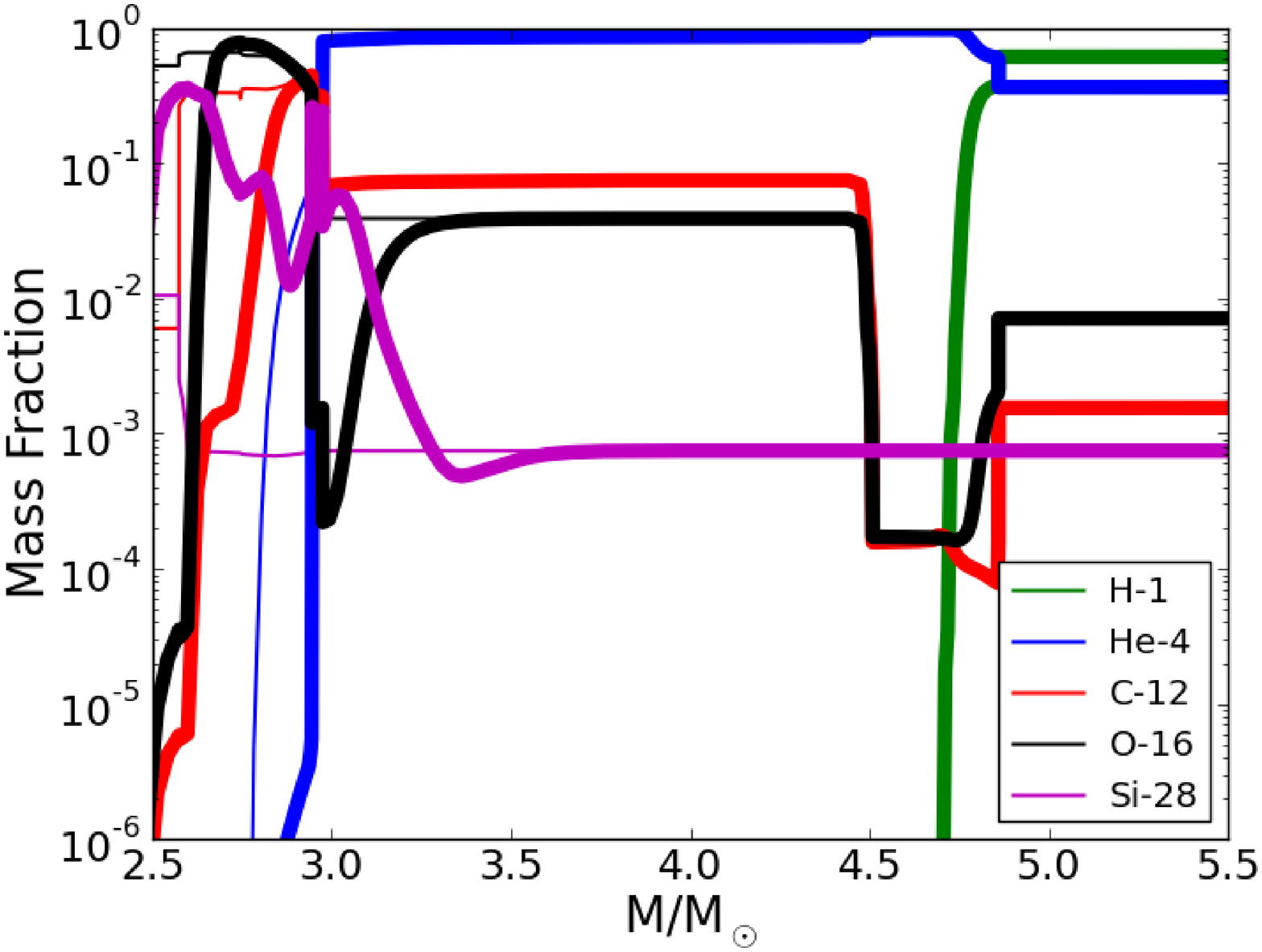}}}
\resizebox{7.5cm}{!}{\rotatebox{0}{\includegraphics{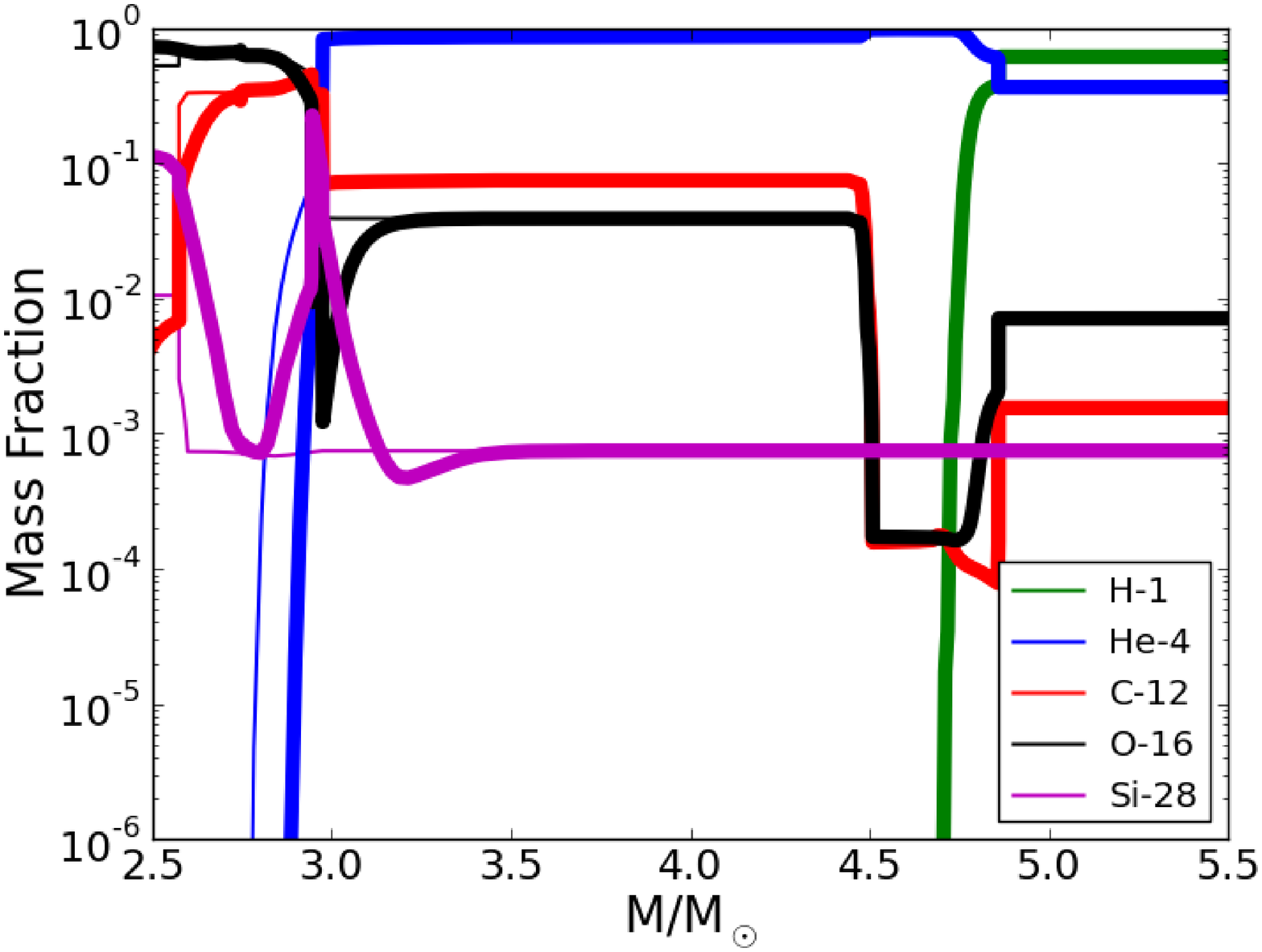}}}
\resizebox{7.5cm}{!}{\rotatebox{0}{\includegraphics{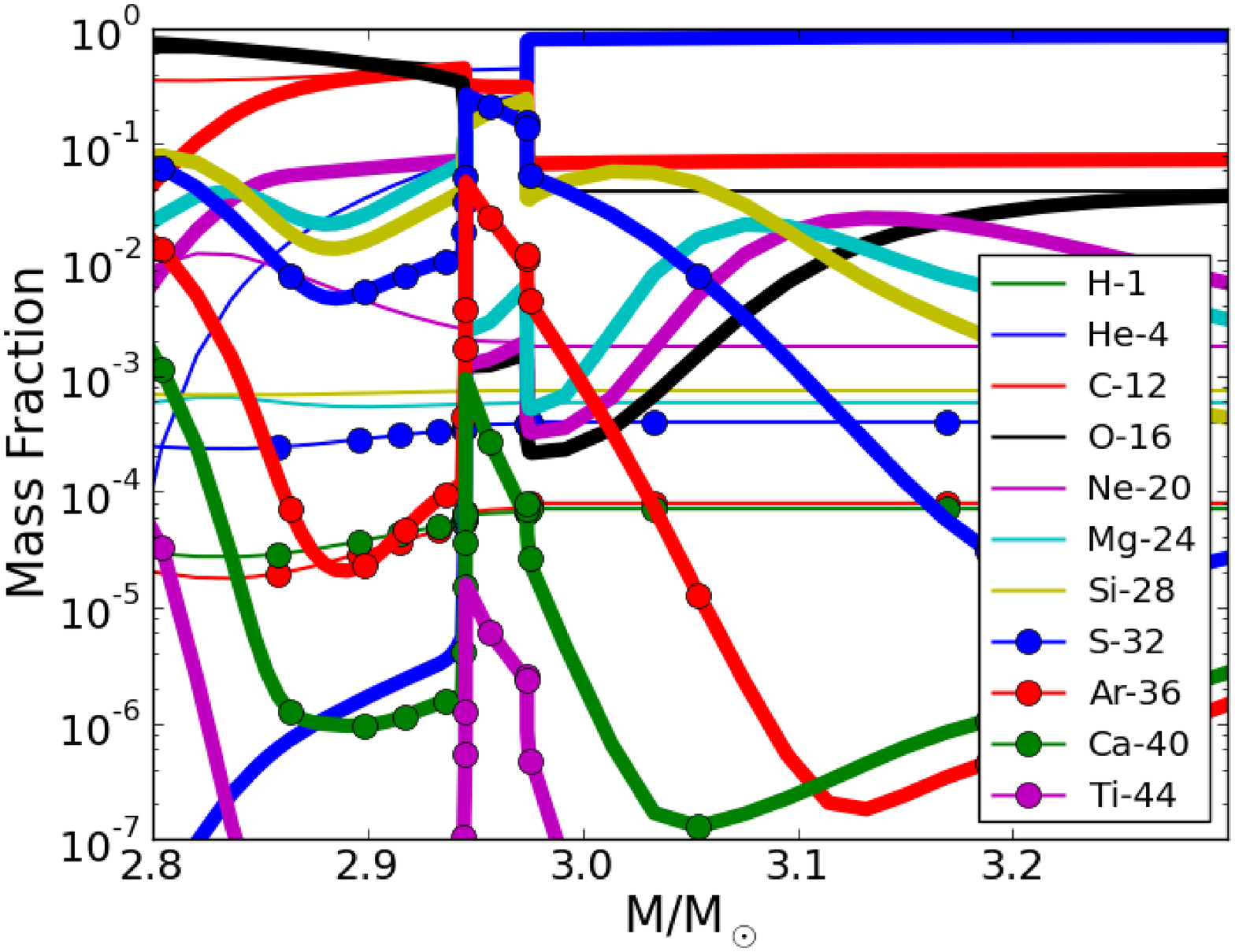}}}
\resizebox{7.5cm}{!}{\rotatebox{0}{\includegraphics{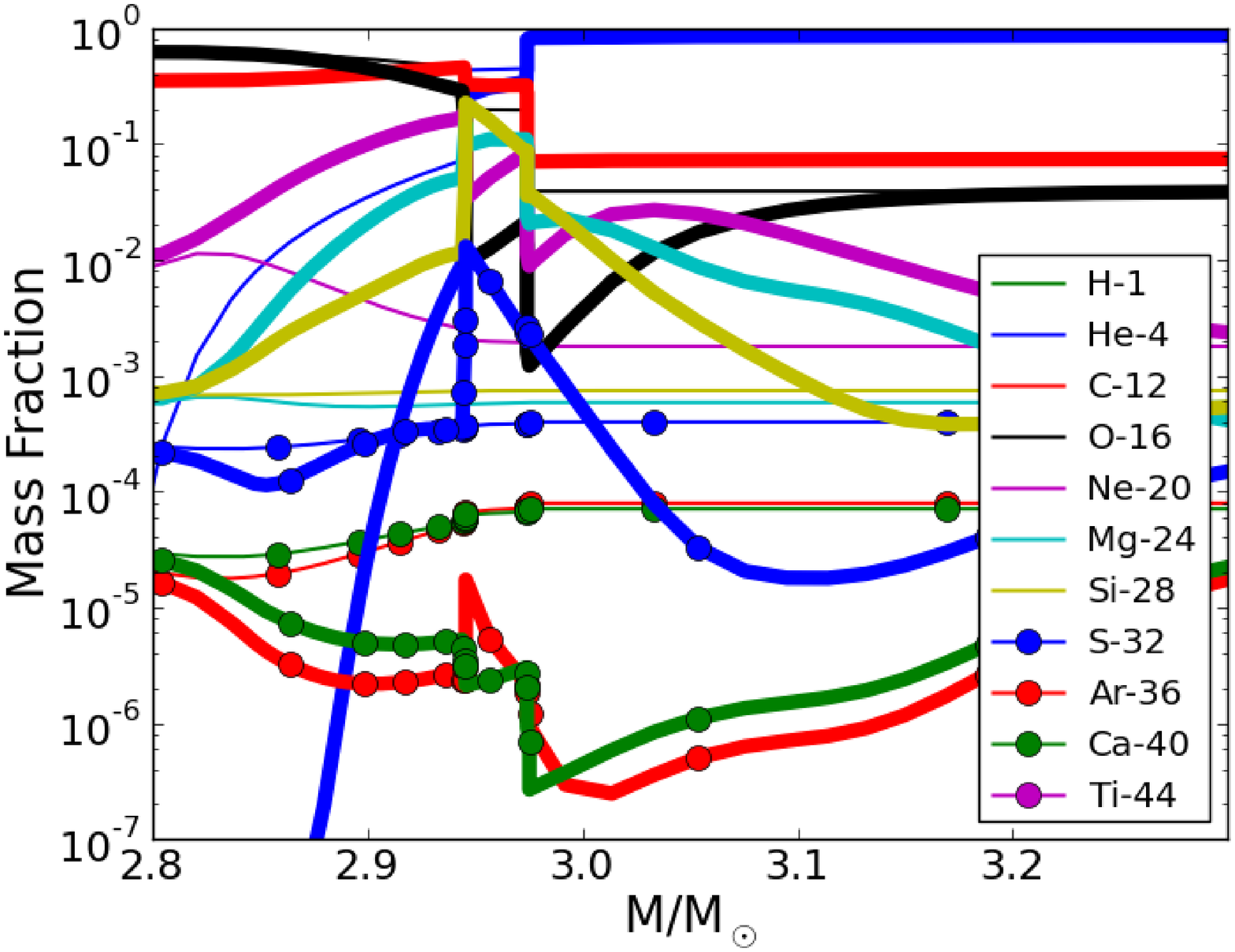}}}
\caption{Abundances profiles in the C-rich material
for H, $^4$He, $^{12}$C, $^{16}$O and
$^{28}$Si before (thin lines) and after (thick lines)
the supernova explosion for models 15r and 15r4
(Upper Left and Right Panels, respectively).
The same is shown in more detail and for more species in the lower panels.
The C/Si zone is where O is depleted producing $^{28}$Si, above
$2.945 \mathrm{M_{\odot}}$.
The process of O-depletion affects a large part of the He/C zone,
up to M = $3.3 \mathrm{M_{\odot}}$ and $3.2 \mathrm{M_{\odot}}$ for models 15r and 15r4,
respectively.
}
\label{fig:15_spagh_he}
\end{figure}

\begin{figure}
\centering
\resizebox{7.5cm}{!}{\rotatebox{0}{\includegraphics{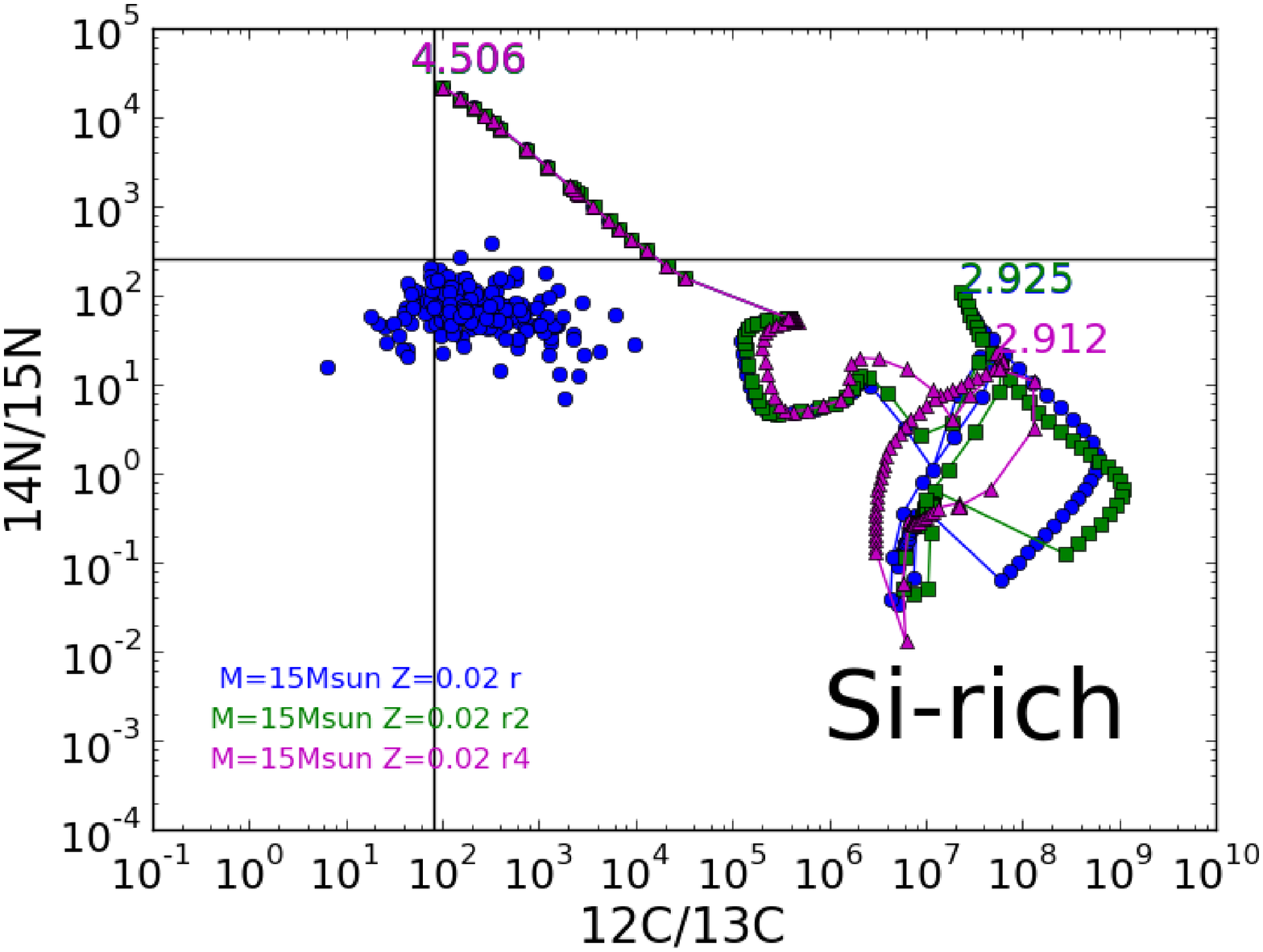}}}
\resizebox{7.5cm}{!}{\rotatebox{0}{\includegraphics{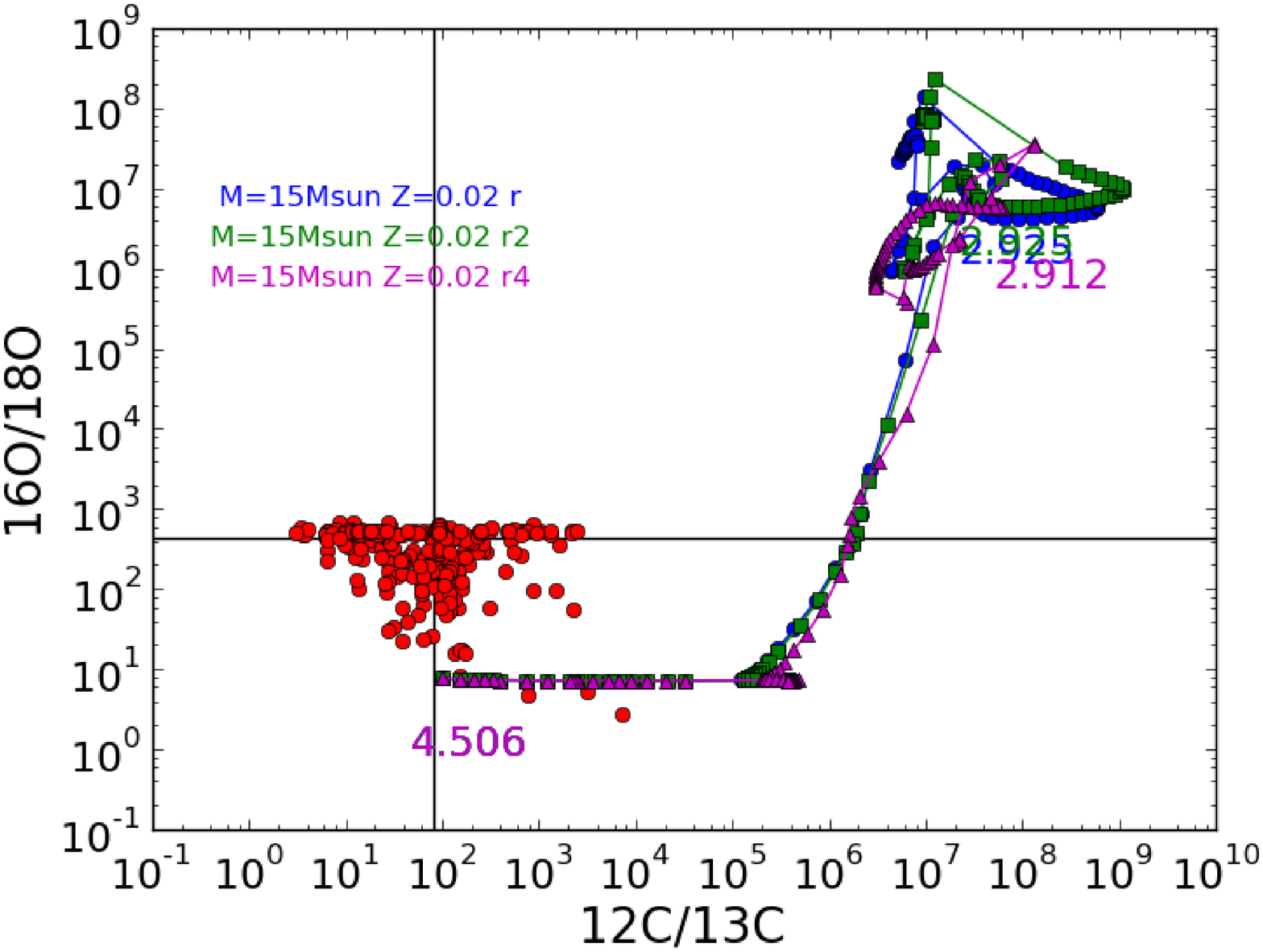}}}
\caption{Isotopic ratios $^{14}$N/$^{15}$N and $^{12}$C/$^{13}$C from
models 15r, 15r2 and 15r4 compared with measured data
for SiC-X grains (Left Panel), and $^{16}$O/$^{18}$O for LD graphites
(Right Panel). The numbers in the plots are mass coordinates where
the C-rich region starts and where it ends in the ejecta.
Different points along the lines indicate different mass-zones where
the nucleosynthesis was calculated.
Continuous black lines indicate the solar ratios.
}
\label{fig:cn_co_ratio}
\end{figure}

In Fig.\,\ref{fig:cn_co_ratio}, Left Panel, C and N isotopic ratios
for SiC-X grains are shown compared to the abundances
in C-rich material from our models. Not included in the figure,
LD graphites show a similar $^{12}$C/$^{13}$C spread compared to SiC-X grains.
The C/Si zone and
the bottom of the He/C zone show high $^{12}$C/$^{13}$C
and low $^{14}$N/$^{15}$N. The carbon isotopic ratio is at least
3 orders of magnitude larger than that of the grains, due to lack of $^{13}$C
(area labeled as "Si-rich" in the plot).
Grains data are located in between our predictions and the solar ratios.
Some mixture with a normal component and/or with
material from the external part of the He shell is required
to fit the observations in this scenario.
Indeed, in the most external part of the He/C zone
(the part that is not O-depleted) the C ratio evolves to normal values,
but with a N ratio larger than solar.
Notice that in Fig.\,\ref{fig:cn_co_ratio}
there is a group of SiC-X grains showing
$^{12}$C/$^{13}$C lower than solar.
We will discuss them in a further work, where SN ejecta affected by
H burning are considered.
In Fig.\,\ref{fig:cn_co_ratio}, we report the comparison
with LD graphites for $^{16}$O/$^{18}$O, and again
$^{12}$C/$^{13}$C.
The largest measured $^{18}$O excess is
observed for few graphites with $^{12}$C/$^{13}$C larger than solar,
in agreement with the abundance signature in the external part of the
He/C zone. In the C/Si zone and at the bottom of the He/C zone
the $^{16}$O/$^{18}$O becomes up to 7-8 orders of magnitude larger
than solar. However, such a large ratio can be reduced
to a ratio $\sim$ solar with minor mixing with normal material,
since there is negligible amount of oxygen left in these layers
(see e.g., Fig.\,\ref{fig:15_spagh_he}).

\begin{figure}
\centering
\resizebox{7.5cm}{!}{\rotatebox{0}{\includegraphics{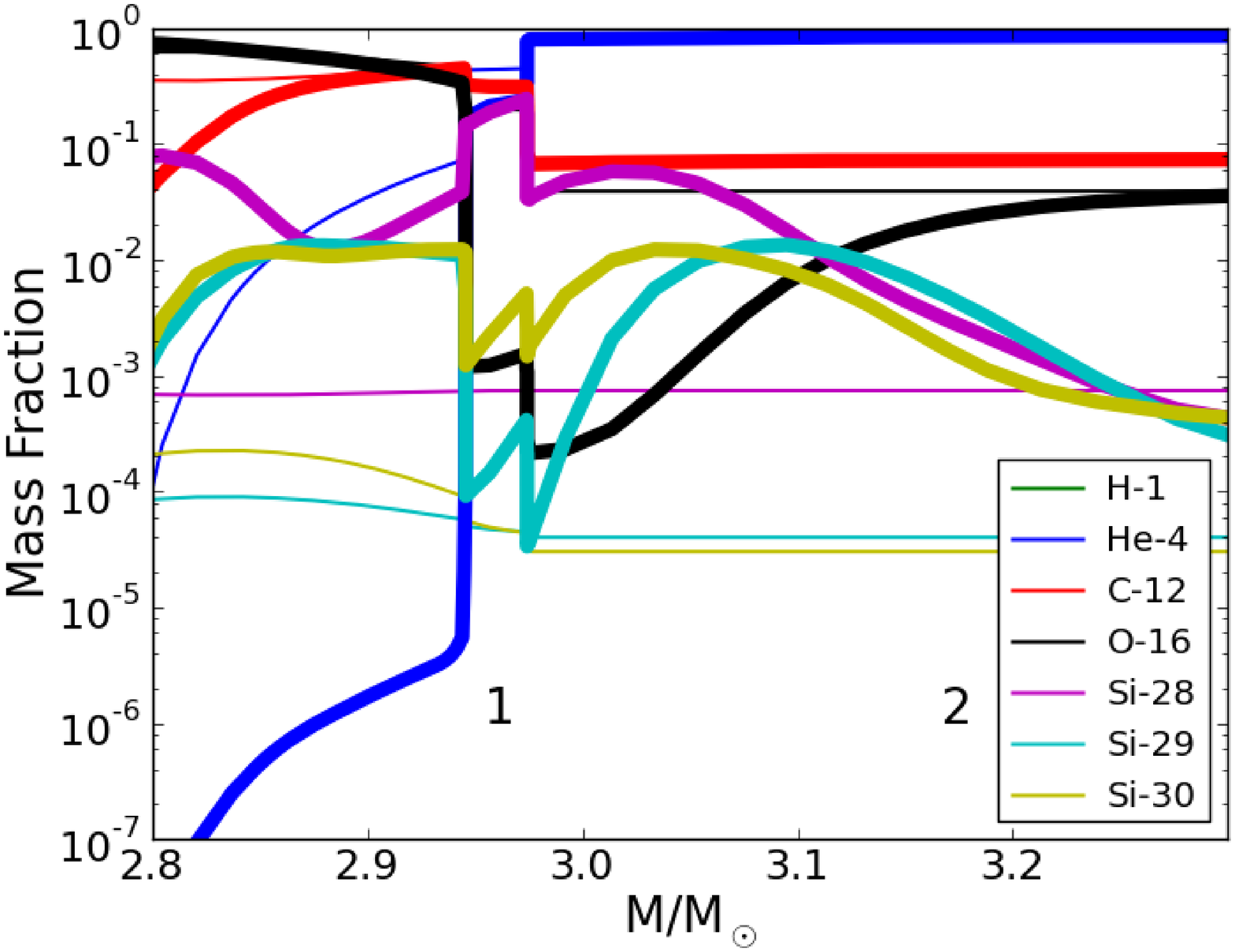}}}
\resizebox{7.5cm}{!}{\rotatebox{0}{\includegraphics{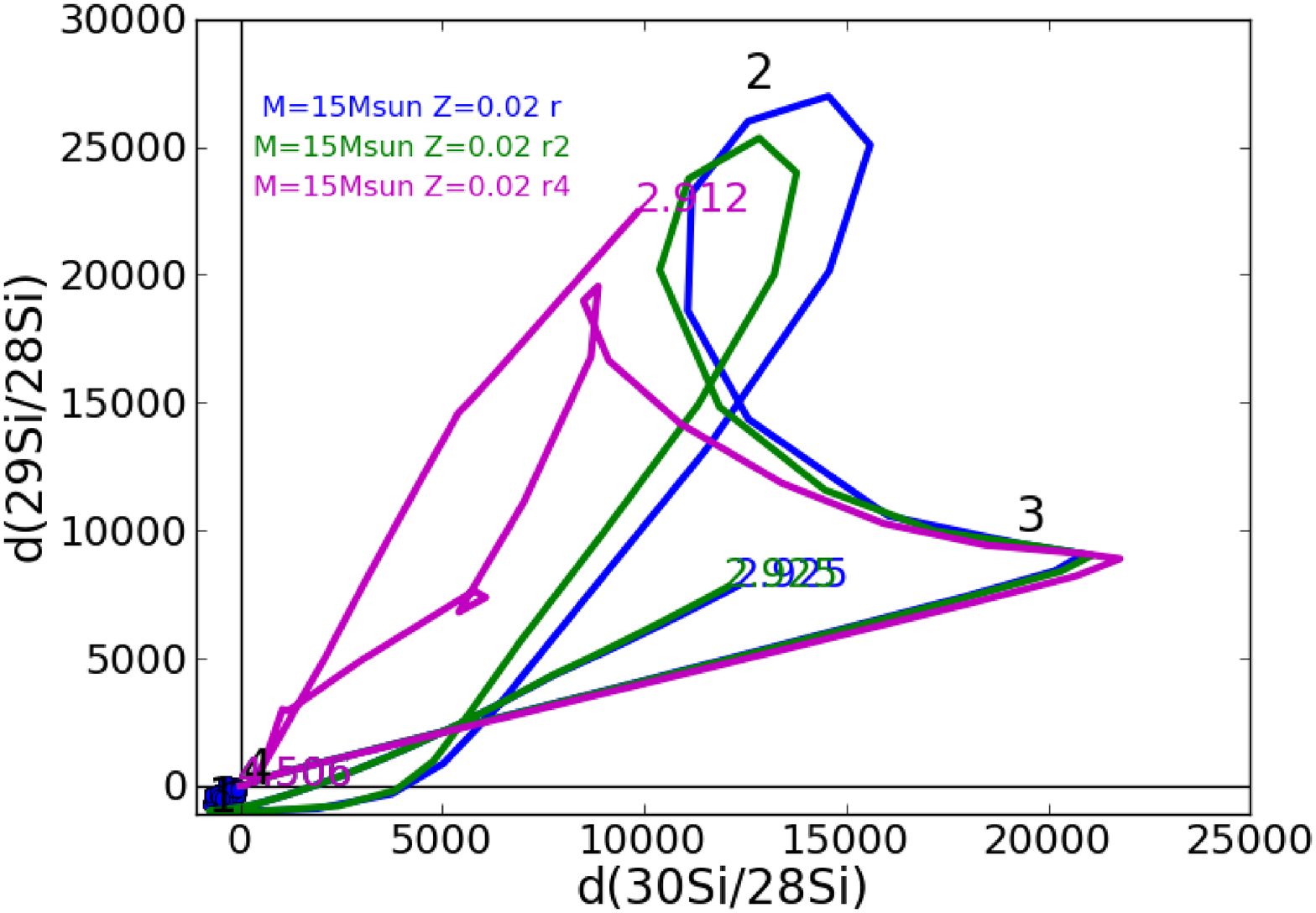}}}
\resizebox{7.5cm}{!}{\rotatebox{0}{\includegraphics{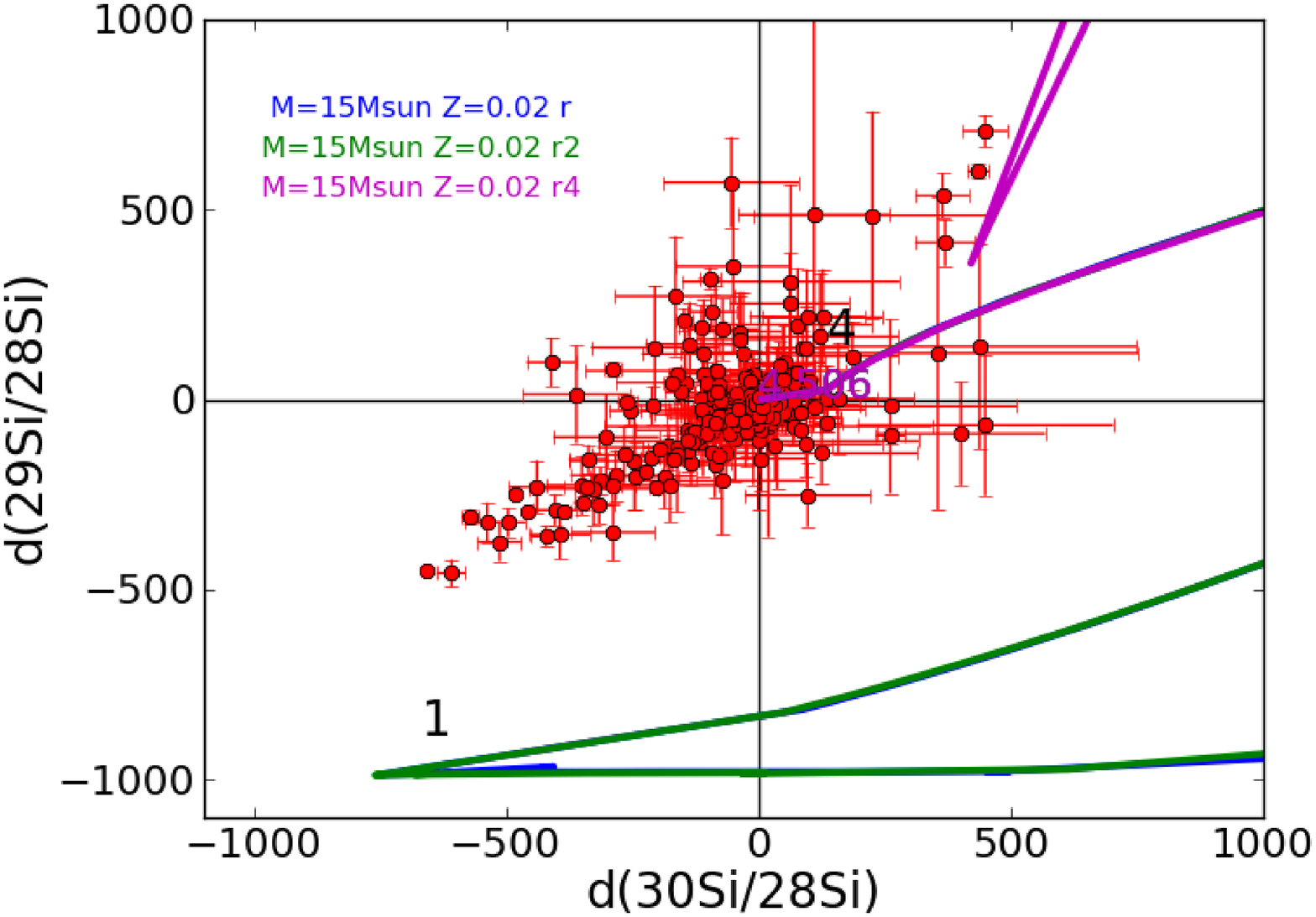}}}
\resizebox{7.5cm}{!}{\rotatebox{0}{\includegraphics{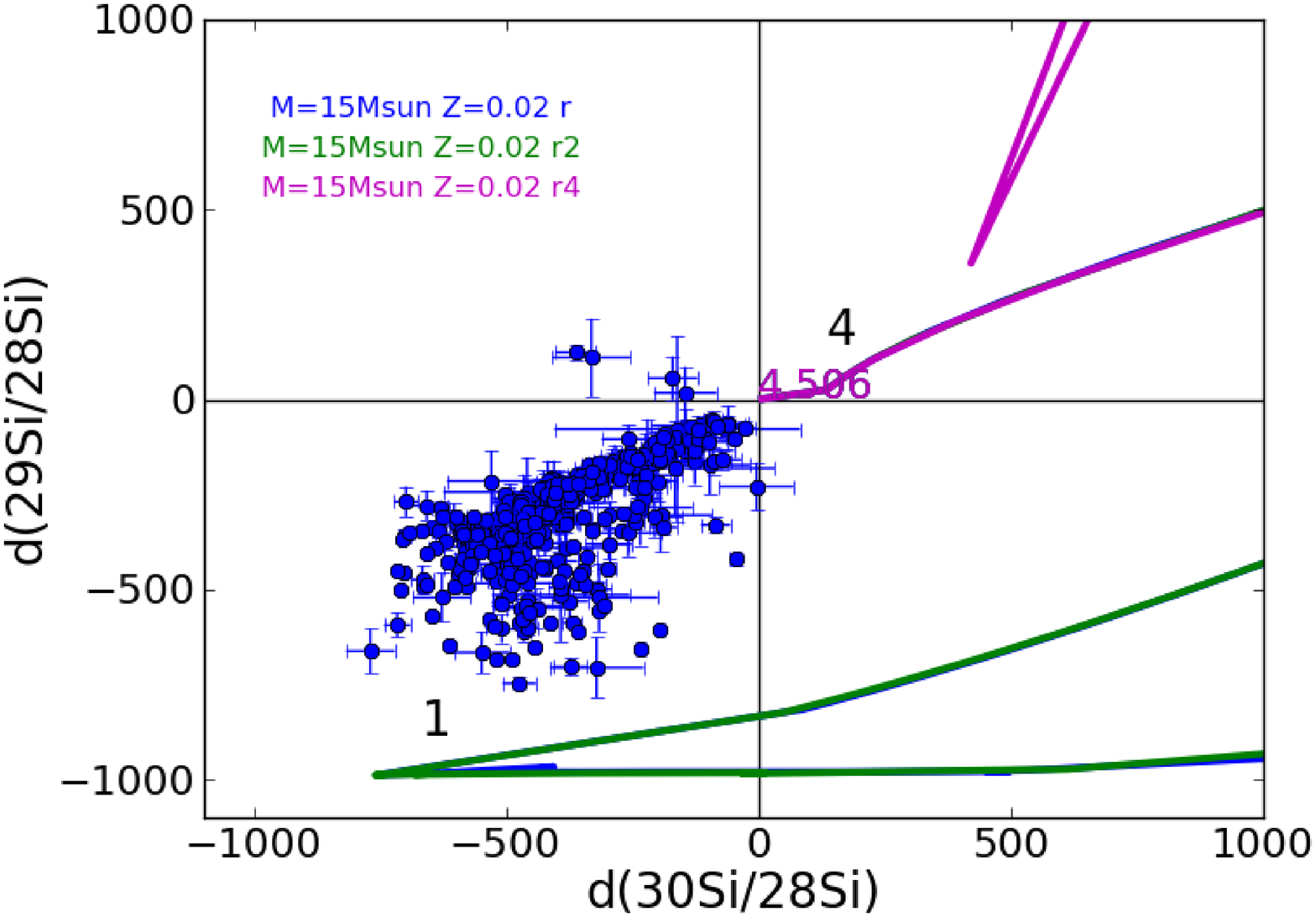}}}
\caption{Si abundances in the Si-rich material for model 15r before and
after CCSN (Upper Left Panel). Si isotopic ratios in $\delta$ notation
($\delta$(ratio) = (stellar ratio/solar ratio - 1)$\times$1000)
is given for SiC-X grains (Upper Right Panel and zoomed in the
Lower Right Panel) and LD graphites (Lower Left Panel) compared with
models. The Si abundances in model 15r and the correspondent $\delta$-values
are highlighted for 4 zones, at M = 2.95, 3.15 (zones $1$ and $2$), 3.3 and
4 M$_{\odot}$ (zones $3$ and $4$, out of the x-axis range in the Upper Left Panel).
}
\label{fig:sidelta_grains}
\end{figure}

In Fig.\,\ref{fig:sidelta_grains}, Right Panels,
the Si isotopic ratio of stellar models and SiC-X grains are compared.
The same is done for LD graphites in the Bottom Left Panel.
In the Upper Left Panel, the pre-explosive and post-explosive
isotopic Si abundances are reported for model 15r.
During the explosion,
in the C/O zone with low $\alpha$-fuel available
$^{28}$Si is marginally produced via $\alpha$-capture, and
$^{29,30}$Si are strongly produced via neutron capture.
The C/Si zone shows a strong $^{28}$Si production.
In particular, in zone $1$ (M = 2.95 M$_{\odot}$)
the pre-explosive temperature and density are $3.1 \times 10^8 \mathrm{K}$
and $1.34 \times 10^3 \mathrm{cm^{-3}}$, rising up to $2.0 \times 10^9 \mathrm{K}$
and $7.58 \times 10^3 \mathrm{cm^{-3}}$ respectively, during the explosion.
Moving outward in the the deepest part of the He/C zone,
the $n$-process \citep[e.g.,][]{meyer:00}
gradually changes the $^{28}$Si-enrichment to strong
$^{29,30}$Si-enrichments.
Less extreme Si isotopic ratios with a mild
$^{29,30}$Si-excess are seen in the external He/C zone,
due to the pre-explosive $s$-processing.
The $^{28}$Si-excess is reproduced in part of the C/Si zone
for models 15r and 15r2 (see e.g. zone $1$ in model 15r, Fig.\,\ref{fig:sidelta_grains},
Upper Left Panel and Right Bottom Panel).
Also in this case, the grains data shows $\delta$-values closer to the solar ratio,
corresponding again to some mixing with unprocessed material, or with more external C-rich
stellar layers.
On the other hand, in model 15r4 the shock temperature is not sufficient to reproduce
the $^{28}$Si-excess observed in SiC-X grains.
The combined nucleosynthesis contribution from $\alpha$-captures and
neutron captures cause a milder $^{28}$Si/$^{30}$Si-enrichment
compared to $^{28}$Si/$^{29}$Si moving outward in the
$^{28}$Si-rich regions, and an earlier positive
$\delta$($^{30}$Si) with decreasing temperature compared to
$\delta$($^{29}$Si).
This trend is consistent with a subset of SiC-X grains,
whereas most of the SiC-X grains belong to the group X1
\citep[][]{lin:02}.

Finally, in Fig.\,\ref{fig:sidelta_grains}, Left Bottom Panel,
LD graphites show a relevant amount of grains with
$^{28}$Si excess, but with
many scattered around the solar ratios, and few with positive Si
$\delta$ ratios. This different behavior between SiC-X grains and
graphites is qualitatively explained looking at
the profile of Si isotopes in model 15r
(Fig.\,\ref{fig:sidelta_grains}, Left Upper Panel).
The efficiency of graphite condensation depends on
the C enrichment only,
whereas SiC-X grains depend on both the C and Si inventory.
Therefore, SiC-X condensation will be more efficient
in the C/Si zone than in the He/C zone,
whereas constraints for graphites are less severe.
We may also expect that the condensation of SiC-X and
graphites will
be less efficient in the external part of the He/C zone,
compared to the C-rich and O-depleted regions. Indeed,
in these stellar layers there is less free C and Si available.
Simulations of the condensation processes
using present abundance predictions are required
to better quantify this analysis.

The initial $^{44}$Ti and $^{26}$Al enrichment are
distinctive signatures for a significant
fraction of SiC-X grains and LD graphites
\citep[e.g.,][]{clayton:04}.
Their initial amount that condenses into the grains is usually estimated assuming
that the $^{44}$Ca/$^{40}$Ca and $^{26}$Mg/$^{24}$Mg ratios are
solar. This is not the case in our calculations (see e.g. $^{40}$Ca abundance profile
in Fig.\,\ref{fig:15_spagh_he}).
Furthermore, Al and Ti
condense more efficiently than Mg and Ca, respectively
\citep[e.g.,][]{amari:95,besmehn:03}.
For model 15r, the
thermodynamic conditions allow to produce $^{40}$Ca and $^{44}$Ti
at the same time, with $^{44}$Ti/$^{40}$Ca $\sim$ 10$^{-2}$
(Fig.\,\ref{fig:15_spagh_he}).
The largest measured $^{44}$Ca/$^{40}$Ca ratio
in SiC-X grains of $\sim$ 2-3 \citep[][]{hynes:09},
would be satisfied with a fractionation Ti/Ca $\sim$ 100
without considering also direct $^{44}$Ca production.
In the He/C zone, the neutron captures deplete the
pre-explosive $^{40}$Ca by orders of magnitude,
eventually leading to high $^{44}$Ca/$^{40}$Ca and high
$^{48}$Ca/$^{40}$Ca. $^{48}$Ti is depleted in this case.
We notice that in measurements
by e.g. \cite{besmehn:03} $^{48}$Ca and $^{48}$Ti were
not possible to distinguish. For model 15r4,
the abundances in the C/Si zone are less extreme than model 15r,
with a nucleosynthesis more similar to the He/C zone.

One of the main puzzles for
the previous ad-hoc mixing scenarios for SiC-X grains
is the non-enhancement of $^{54}$Fe
compared to $^{56}$Fe, since in standard CCSN
models $^{28}$Si-rich zones are also $^{54}$Fe-rich.
In the C/Si zone of our models Fe species
are destroyed by neutron captures and grain signatures
with low $^{54}$Fe are predicted.
The isotopes $^{57,58}$Fe can also be produced outward
in the He/C zone.
%
%

\section{Conclusions and final remarks}
\label{sec: summary}

We have compared the abundance signature in presolar
SiC-X grains and LD graphites with nucleosynthesis predictions
from CCSN ejecta exposed to high shock velocities.
Due to basic nuclear properties, an $\alpha$-capture chain
starts from $^{16}$O feeding heavier species including $^{28}$Si,
without affecting the pre-explosive abundance of $^{12}$C
and creating a new C/Si zone.
The amount and distribution of $\alpha$-isotopes
mainly depend on the abundance of the seed $^{16}$O,
on the explosion shock velocity and on the amount of $\alpha$-particles
available.
The interplay between the $\alpha$-captures and the
$n$-process regulates the relative abundances of $^{28,29,30}$Si
in the He/C zone.

The typical isotopic trends for C, N, and O of LD graphites and SiC-X
grains can be reproduced using the C-rich ejecta,
with the exception of grains with low
$^{12}$C/$^{13}$C.
%
The observed $^{28}$Si-excess is explained
qualitatively from the signature in the C/Si zone, where
$^{16}$O has been depleted by following $\alpha$-captures
to build $^{28}$Si and other heavier species.
Since production of graphites depends only on the C
abundance, and SiC-X on both C and Si, this may explain
why the $^{28}$Si-excess is not observed in all LD
graphites. Therefore, the C/Si zone is an ideal environment
to condense SiC-X and LD graphites.
Such abundance signatures triggered
by high temperatures can be also obtained in parts of
asymmetric CCSN ejecta exposed to high shock velocities, or in
hypernovae ejecta.

We reconsidered the $^{44}$Ti-excess extrapolated from a sample of
C-rich presolar grains. In the C/Si zone the
$^{44}$Ca/$^{40}$Ca is not solar, affecting the results of such
extrapolation.
The Ca and Ti condensation
efficiency is uncertain, leading to their elemental fractionation.
For the model with largest explosion energy (model 15r)
both $^{40}$Ca and $^{44}$Ti (and $^{48}$Ti) are produced in the
C/Si zone (with $^{40}$Ca/$^{44}$Ti $\sim$ 100).
Neutron captures destroy
$^{40}$Ca in the He/C zone. The isotope $^{44}$Ca (and $^{48}$Ca)
instead is produced by neutron
captures in all the O-depleted and C-rich ejecta.
For weaker shocks (see e.g., model 15r4)
neutron captures dominate the Ca nucleosynthesis,
$^{40}$Ca is depleted and $^{44}$Ca produced.
In both cases, a comparison needs to be made grain-by-grain,
where correlations with e.g., Ti species and $^{28}$Si-excess
\citep[e.g.,][]{clayton:04} are considered.
A similar discussion can be derived for $^{26}$Al enrichment.

Finally, the present results provide support to the existence
of different energy components in the ejecta of the same CCSN.
Asymmetries during the SN explosion of e.g., SN1987A and
CasA are confirmed by observations.
In SN1987A the emission SiI and FeI-II from
core material are correlated with HeI emission, possibly indicating
the signature of $\alpha$-rich freezout during high-energy-SN explosion
in at least part of the ejecta \citep[][]{kiaer:10}.
In CasA, \cite{delaney:10} reported two major velocity
components for SiII in the ejecta. We suggest that
they may be the signature of the Si/S-Si/O and C/Si zones,
respectively. This scenario would also explain
the non-correlation between NeII and OIV emission in CasA
\citep[][]{isensee:10}. Indeed, Ne
(together with Mg, S, Ar and Ca) could be produced
also in the C/Si zone, whereas for standard CCSN
O and Ne are produced in the same regions.

\acknowledgments
NuGrid acknowledges significant support from NSF grants PHY 02-16783
and PHY 09-22648 (Joint Institute for Nuclear Astrophysics, JINA) and
EU MIRG-CT-2006-046520. The continued work on codes and in disseminating
data is made possible through funding from STFC and EU-FP7-ERC-2012-St
Grant 306901 (RH, UK), and NSERC
Discovery grant (FH, Canada), and an Ambizione grant of the SNSF
(MP, Switzerland). MP also thanks support from EuroGENESIS.
NuGrid data is served by Canfar/CADC. We thank the anonymous
reviewer for detailed comments and suggestions that greatly
improved the manuscript.

\hyphenation{Post-Script Sprin-ger}

\clearpage

%

\begin{thebibliography}{31}
\expandafter\ifx\csname natexlab\endcsname\relax\def\natexlab#1{#1}\fi

\bibitem[{{Amari} {et~al.}(1992){Amari}, {Hoppe}, {Zinner}, \&
  {Lewis}}]{amari:92}
{Amari}, S., {Hoppe}, P., {Zinner}, E., \& {Lewis}, R.~S. 1992, \apjl, 394, L43

\bibitem[{{Amari} {et~al.}(1995){Amari}, {Lewis}, \& {Anders}}]{amari:95}
{Amari}, S., {Lewis}, R.~S., \& {Anders}, E. 1995, \gca, 59, 1411

\bibitem[{{Bennett} {et~al.}(2012){Bennett}, {Hirschi}, {Pignatari}, {Diehl},
  {Fryer}, {Herwig}, {Hungerford}, {Nomoto}, {Rockefeller}, {Timmes}, \&
  {Wiescher}}]{bennett:12}
{Bennett}, M.~E., {et~al.} 2012, \mnras, 420, 3047

\bibitem[{{Besmehn} \& {Hoppe}(2003)}]{besmehn:03}
{Besmehn}, A., \& {Hoppe}, P. 2003, \gca, 67, 4693

\bibitem[{{Blake} \& {Schramm}(1976)}]{blake:76}
{Blake}, J.~B., \& {Schramm}, D.~N. 1976, \apj, 209, 846

\bibitem[{{Buchmann} \& {Barnes}(2006)}]{buchmann:06}
{Buchmann}, L.~R., \& {Barnes}, C.~A. 2006, Nuclear Physics A, 777, 254

\bibitem[{{Caughlan} {et~al.}(1985){Caughlan}, {Fowler}, {Harris}, \&
  {Zimmerman}}]{caughlan:85}
{Caughlan}, G.~R., {Fowler}, W.~A., {Harris}, M.~J., \& {Zimmerman}, B.~A.
  1985, Atomic Data and Nuclear Data Tables, 32, 197

\bibitem[{{Clayton}(2013)}]{clayton:13}
{Clayton}, D.~D. 2013, \apj, 762, 5

\bibitem[{{Clayton} \& {Nittler}(2004)}]{clayton:04}
{Clayton}, D.~D., \& {Nittler}, L.~R. 2004, \araa, 42, 39

\bibitem[{{Costantini} {et~al.}(2010){Costantini}, {Deboer}, {Azuma}, {Couder},
  {G{\"o}rres}, {Hammer}, {Leblanc}, {Lee}, {O'Brien}, {Palumbo}, {Simpson},
  {Stech}, {Tan}, {Uberseder}, \& {Wiescher}}]{costantini:10}
{Costantini}, H., {et~al.} 2010, \prc, 82, 035802

\bibitem[{{DeLaney} {et~al.}(2010){DeLaney}, {Rudnick}, {Stage}, {Smith},
  {Isensee}, {Rho}, {Allen}, {Gomez}, {Kozasa}, {Reach}, {Davis}, \&
  {Houck}}]{delaney:10}
{DeLaney}, T., {et~al.} 2010, \apj, 725, 2038

\bibitem[{{Fryer} {et~al.}(2012){Fryer}, {Belczynski}, {Wiktorowicz},
  {Dominik}, {Kalogera}, \& {Holz}}]{fryer:12}
{Fryer}, C.~L., {Belczynski}, K., {Wiktorowicz}, G., {Dominik}, M., {Kalogera},
  V., \& {Holz}, D.~E. 2012, \apj, 749, 91

\bibitem[{{Hynes} \& {Gyngard}(2009)}]{hynes:09}
{Hynes}, K.~M., \& {Gyngard}, F. 2009, in Lunar and Planetary Institute Science
  Conference Abstracts, Vol.~40, Lunar and Planetary Institute Science
  Conference Abstracts, 1198

\bibitem[{{Isensee} {et~al.}(2010){Isensee}, {Rudnick}, {DeLaney}, {Smith},
  {Rho}, {Reach}, {Kozasa}, \& {Gomez}}]{isensee:10}
{Isensee}, K., {Rudnick}, L., {DeLaney}, T., {Smith}, J.~D., {Rho}, J.,
  {Reach}, W.~T., {Kozasa}, T., \& {Gomez}, H. 2010, \apj, 725, 2059

\bibitem[{{Kj{\ae}r} {et~al.}(2010){Kj{\ae}r}, {Leibundgut}, {Fransson},
  {Jerkstrand}, \& {Spyromilio}}]{kiaer:10}
{Kj{\ae}r}, K., {Leibundgut}, B., {Fransson}, C., {Jerkstrand}, A., \&
  {Spyromilio}, J. 2010, \aap, 517, A51

\bibitem[{{Lin} {et~al.}(2002){Lin}, {Amari}, \& {Pravdivtseva}}]{lin:02}
{Lin}, Y., {Amari}, S., \& {Pravdivtseva}, O. 2002, \apj, 575, 257

\bibitem[{{Marhas} {et~al.}(2008){Marhas}, {Amari}, {Gyngard}, {Zinner}, \&
  {Gallino}}]{marhas:08}
{Marhas}, K.~K., {Amari}, S., {Gyngard}, F., {Zinner}, E., \& {Gallino}, R.
  2008, \apj, 689, 622

\bibitem[{{Meakin} \& {Arnett}(2007)}]{meakin:07}
{Meakin}, C.~A., \& {Arnett}, D. 2007, \apj, 667, 448

\bibitem[{{Meyer} {et~al.}(2000){Meyer}, {Clayton}, \& {The}}]{meyer:00}
{Meyer}, B.~S., {Clayton}, D.~D., \& {The}, L.-S. 2000, \apjl, 540, L49

\bibitem[{{Meyer} {et~al.}(1995){Meyer}, {Weaver}, \& {Woosley}}]{meyer:95}
{Meyer}, B.~S., {Weaver}, T.~A., \& {Woosley}, S.~E. 1995, Meteoritics, 30, 325

\bibitem[{{Meynet} {et~al.}(2006){Meynet}, {Ekstr{\"o}m}, \&
  {Maeder}}]{meynet:06}
{Meynet}, G., {Ekstr{\"o}m}, S., \& {Maeder}, A. 2006, \aap, 447, 623

\bibitem[{{Nittler} \& {Alexander}(2002)}]{nittler:02}
{Nittler}, L.~R., \& {Alexander}, C.~M.~O. 2002, Meteoritics and Planetary
  Science Supplement, 37, 110

\bibitem[{{Nomoto} {et~al.}(2009){Nomoto}, {Wanajo}, {Kamiya}, {Tominaga}, \&
  {Umeda}}]{nomoto:09}
{Nomoto}, K., {Wanajo}, S., {Kamiya}, Y., {Tominaga}, N., \& {Umeda}, H. 2009,
  in IAU Symposium, Vol. 254, IAU Symposium, ed. J.~{Andersen},
  J.~{Bland-Hawthorn}, \& B.~{Nordstr{\"o}m}, 355--368

\bibitem[{{Rauscher} {et~al.}(2002){Rauscher}, {Heger}, {Hoffman}, \&
  {Woosley}}]{rauscher:02}
{Rauscher}, T., {Heger}, A., {Hoffman}, R.~D., \& {Woosley}, S.~E. 2002, \apj,
  576, 323

\bibitem[{{Schmalbrock} {et~al.}(1983){Schmalbrock}, {Becker}, {Buchmann},
  {G{\"o}rres}, {Kettner}, {Kieser}, {Kr{\"a}winkel}, {Rolfs}, {Trautvetter},
  {Hammer}, \& {Azuma}}]{schmalbrock:83}
{Schmalbrock}, P., {et~al.} 1983, Nuclear Physics A, 398, 279

\bibitem[{{Strandberg} {et~al.}(2008){Strandberg}, {Beard}, {Couder},
  {Couture}, {Falahat}, {G{\"o}rres}, {Leblanc}, {Lee}, {O'Brien}, {Palumbo},
  {Stech}, {Tan}, {Ugalde}, {Wiescher}, {Costantini}, {Scheller}, {Pignatari},
  {Azuma}, \& {Buchmann}}]{strandberg:08}
{Strandberg}, E., {et~al.} 2008, \prc, 77, 055801

\bibitem[{{Thielemann} {et~al.}(1996){Thielemann}, {Nomoto}, \&
  {Hashimoto}}]{thielemann:96}
{Thielemann}, F.-K., {Nomoto}, K., \& {Hashimoto}, M.-A. 1996, \apj, 460, 408

\bibitem[{{Travaglio} {et~al.}(1999){Travaglio}, {Gallino}, {Amari}, {Zinner},
  {Woosley}, \& {Lewis}}]{travaglio:99}
{Travaglio}, C., {Gallino}, R., {Amari}, S., {Zinner}, E., {Woosley}, S., \&
  {Lewis}, R.~S. 1999, \apj, 510, 325

\bibitem[{{Woosley} \& {Weaver}(1995)}]{woosley:95}
{Woosley}, S.~E., \& {Weaver}, T.~A. 1995, \apjs, 101, 181

\bibitem[{{Yoshida}(2006)}]{yoshida:06}
{Yoshida}, T. 2006, \nar, 50, 600

\bibitem[{{Zinner}(2003)}]{zinner:03}
{Zinner}, E.~K. 2003, Treatise on Geochemistry, 1, 17

\end{thebibliography}


\newpage


\end{document}